\documentclass{PoS}

\def\jcap{JCAP}%
\def\prl{Phys.\ Rev.\ Lett.}%

\usepackage{booktabs}
\usepackage{xspace}
\usepackage{multirow}

\usepackage[
  tbtags
]{amsmath}
\usepackage[
  product-units = single,
  output-product = \times,
  inter-unit-product = \cdot
]{siunitx}

\newcommand{\GB}{\textsf{GAMBIT}\xspace}
\newcommand{\Diver}{\textsf{Diver}\xspace}
\newcommand{\MultiNest}{\textsf{MultiNest}\xspace}
\newcommand{\multinest}{\textsf{MultiNest}\xspace}
\newcommand{\flexiblesusy}{\textsf{FlexibleSUSY}\xspace}

\newcommand{\DarkSUSY}{\textsf{DarkSUSY}\xspace}
\newcommand{\DDCalc}{\textsf{DDCalc}\xspace}

\newcommand{\nuLike}{\textsf{nulike}\xspace}
\newcommand\great{\textsf{GreAT}\xspace}
\newcommand\twalk{\textsf{T-Walk}\xspace}
\newcommand\diver{\textsf{Diver}\xspace}

\newcommand{\sss}{\scriptscriptstyle}
\newcommand{\ms}{m_{\sss S}}
\newcommand{\lhs}{\lambda_{h\sss S}}

\newcommand{\mh}{m_h}

\newcommand{\scannerbit}{\textsf{ScannerBit}\xspace}

\title{Global fits of the scalar singlet model using GAMBIT}

\ShortTitle{Global fits of the scalar singlet model using GAMBIT}

\author{\speaker{James McKay} {\rm on behalf of the GAMBIT collaboration}\\
 Department of Physics, Imperial College London, Blackett Laboratory, Prince Consort Road, London SW7 2AZ, UK\\
        E-mail: \email{j.mckay14@imperial.ac.uk}}

\abstract{The extension of the standard model (SM) by a Higgs portal scalar field is one the simplest dark matter theories.  We present here the first results for a global fit to this model using the global and beyond the SM inference tool (GAMBIT).  This software enables the combination of dark matter constraints in a statistically consistent manner.  In total 15 parameters are varied and the parameter space explored using four different scanning algorithms.  The viable parameter space is reduced from previous studies of this model due to the inclusion of the latest direct detection constraints.}

\FullConference{The European Physical Society Conference on High Energy Physics\\
		5-12 July, 2017\\
		Venice}

\begin{document}

\section{Introduction}

There are a multitude of both experimental and theoretical constraints on the nature of physics beyond the standard model (SM).  For any beyond SM theory, the allowed values of its parameters are determined by such constraints.  Combing as many of these as possible in a statistically consistent manner is the process of a \textit{global fit}.  The \GB software \cite{Athron:2017ard,Workgroup:2017htr,Workgroup:2017bkh,Workgroup:2017lvb,Balazs:2017moi,Workgroup:2017myk} was developed to perform global fits in the most flexible and modular way, enabling new models and constraints to be included as efficiently as possible.

The results from the first global fit using \GB applied to the scalar singlet dark matter model are summarised here, with the complete study presented elsewhere \cite{Athron:2017kgt}.  \GB has also been used for global fits of the minimal supersymmetric standard model (MSSM).  In particular in the constrained MSSM (CMSSM) \cite{Athron:2017qdc}, two variants of the non-universal Higgs mass (NUHM) model \cite{Athron:2017qdc} and the MSSM with parameters defined at the weak scale \cite{Athron:2017yua}.

\GB enables the user to incorporate existing software via a backend system.  The following external codes were used to produce the results presented here:  \Diver \cite{Workgroup:2017htr}, \MultiNest 3.10 \cite{Feroz:2008xx} and \great \cite{great} (efficient sampling); \flexiblesusy 1.5.1 \cite{Athron:2014yba} (vacuum stability calculation);  \DDCalc 1.0.0 \cite{Workgroup:2017lvb} (direct detection), \nuLike 1.0.4 \cite{Scott:2012mq,IC79_SUSY} (neutrino indirect detection), gamLike 1.0.0 \cite{Workgroup:2017lvb} (gamma-ray indirect detection) and \DarkSUSY 5.1.3 \cite{Gondolo:2004sc} (Boltzmann solver).  Input files, samples and best-fit benchmarks for this study are publicly accessible from \textsf{Zenodo} \cite{the_gambit_collaboration_2017_801511}.

\section{Model, constraints and scan details}

Scalar singlet dark matter is one of the simplest extensions of the SM.  One real singlet scalar $S$ with mass $\ms$, which is stabilised by a $\mathbb{Z}_2$ symmetry, is coupled to the SM via the Higgs field with a \textit{portal} coupling $\lhs$.  In addition to $S$ being a dark matter candidate, due to the coupling with the Higgs field it also stabilises the electroweak Higgs vacuum for values of $\lhs$ large enough to affect the running of the Higgs coupling.  In general a quartic $S$ coupling is also possible, but has very little phenomenological relevance for the constraints we consider here, so we set it to zero. Thus we have two beyond SM parameters, $\lhs$ and $\ms$.

In addition to these we vary 13 `nuisance' parameters.  These are: the local dark matter (DM) density, the nuclear matrix elements (strange, and up + down), the strong, electromagnetic and Fermi couplings, the Higgs mass, and the six quark masses.  In our scans we allow for these parameters to vary from their central values by at least $\pm3\sigma$ --  see Table 2 in Ref. \cite{Athron:2017kgt} for the central values and allowed ranges.

The combined likelihood is composed of 19 seperate likelihood functions.  Experimental DM constraints include direct detection from LUX Run I (2015) and II (2016)~\cite{LUX2016,LUXrun2}, PandaX (2016)~\cite{PandaX2016}, SuperCDMS (2014) \cite{SuperCDMS}  and XENON100 (2012)~\cite{XENON2013}.  We also constrain the dark matter annihilation cross-section using the IceCube 79 string analysis \cite{IC79,IC79_SUSY} of solar neutrinos and the lack of anomalous gamma ray emission from dwarf spheroidal galaxies from the \textit{Fermi}-LAT experiment~\cite{LATdwarfP8}.  The decay width of the Higgs to invisible (DM) particles is constrained by the non-observation of this process at the Large Hadron Collider.

The thermal relic abundance of scalar particles, $\Omega_S$, is constrained by the \textit{Planck} \cite{Planck15cosmo} measured value ($\Omega_\text{DM} h^2 = 0.1188\pm 0.0010$), which we apply as an upper limit likelihood as described in Sec 8.3.4 of Ref.\ \cite{Athron:2017ard}.  For consistency, if the model under-populates the DM relic density, we rescale all direct and indirect signals to account for the fraction of DM that is detectable.  This is a conservative approach as it suppresses direct and indirect signals in regions where the thermal abundance is less than the \textit{Planck} value.  Finally, via our choice of prior, we place a theoretical constraint of $\lhs\leq10$ from the consideration of perturbative unitarity.  The remaining 10 constraints are simple likelihoods for the nuisance parameters.

We carry out two main types of scan: one over the full range of scalar masses from 45\,GeV to 10\,TeV, intended to sample the entire accessible parameter space, and another centred on lower masses at and below the Higgs resonance $\ms\sim \mh/2$, in order to obtain a more detailed picture of this region.  Four scanning algorithms interfaced via \scannerbit \cite{Workgroup:2017htr} are used for efficient sampling of the 15 dimensional parameter space.  These are a differential evolution sampler \diver \cite{Workgroup:2017htr}, an ensemble Markov Chain Monte Carlo (MCMC) known as \twalk \cite{Workgroup:2017htr}, the \multinest nested sampling algorithm \cite{Feroz:2008xx} and an MCMC implementation via the \great \cite{great} package.  This choice of scanners allows efficient sampling of the multimodal parameter space, providing state of the art optimisation, sampling of the profile likelihood and accurate calculation of the Bayesian posterior.

We performed a total of nine scans, consisting of a low and high mass scan for each of the four scanners, plus one additional scan with \diver focused specifically on the $\ms\sim \mh/2$ region.  These combined to give a total of $5.7\times10^7$ valid samples.  In addition we carried out over 200 more scans as part of scanner performance testing, using the same combined likelihood, with varying numbers of free parameters and scanner settings (such as convergence tolerance) in Ref. \cite{Workgroup:2017htr}.  Based on these results we choose the most stringent scanner settings for the nine scans in this global fit.

\section{Results}
\begin{figure*}[h!]
\centering
\includegraphics[height=0.45\columnwidth]{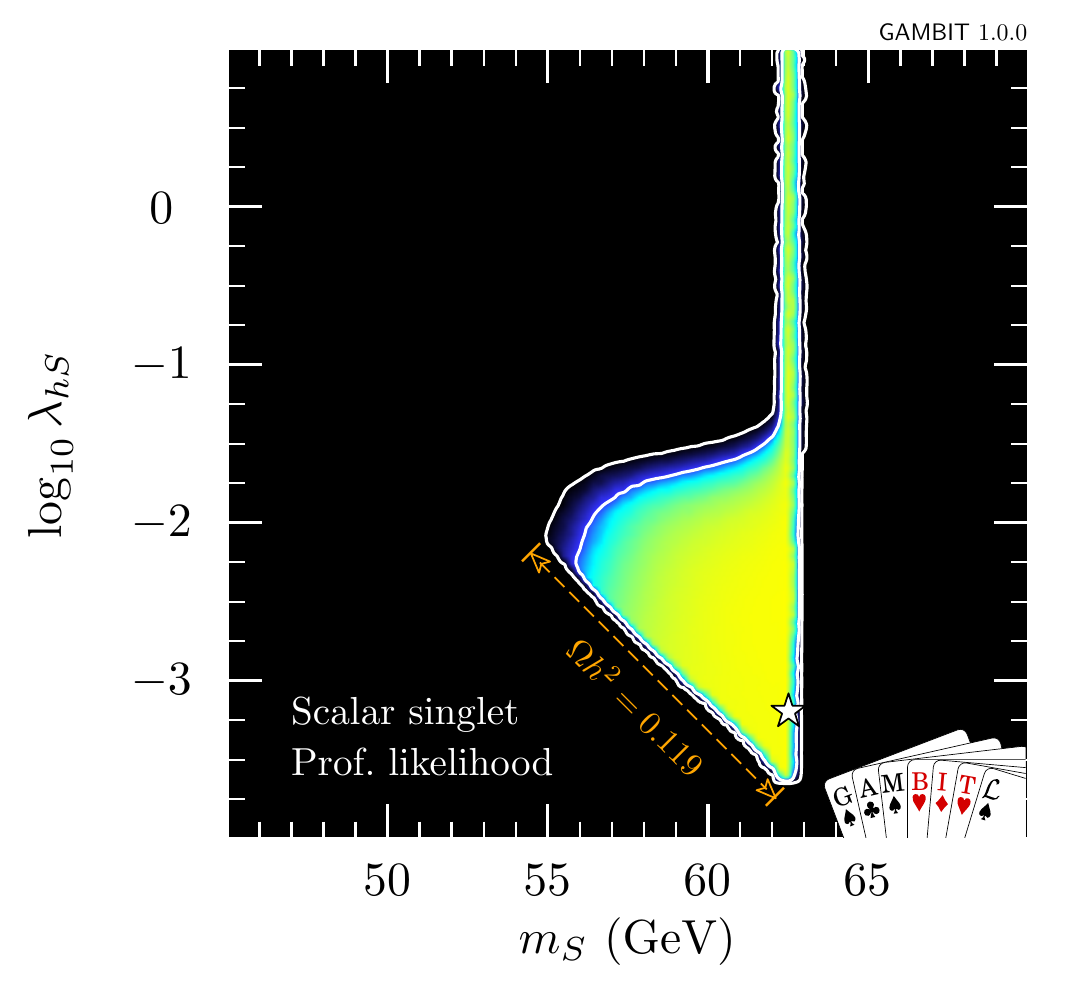}\includegraphics[height=0.45\columnwidth]{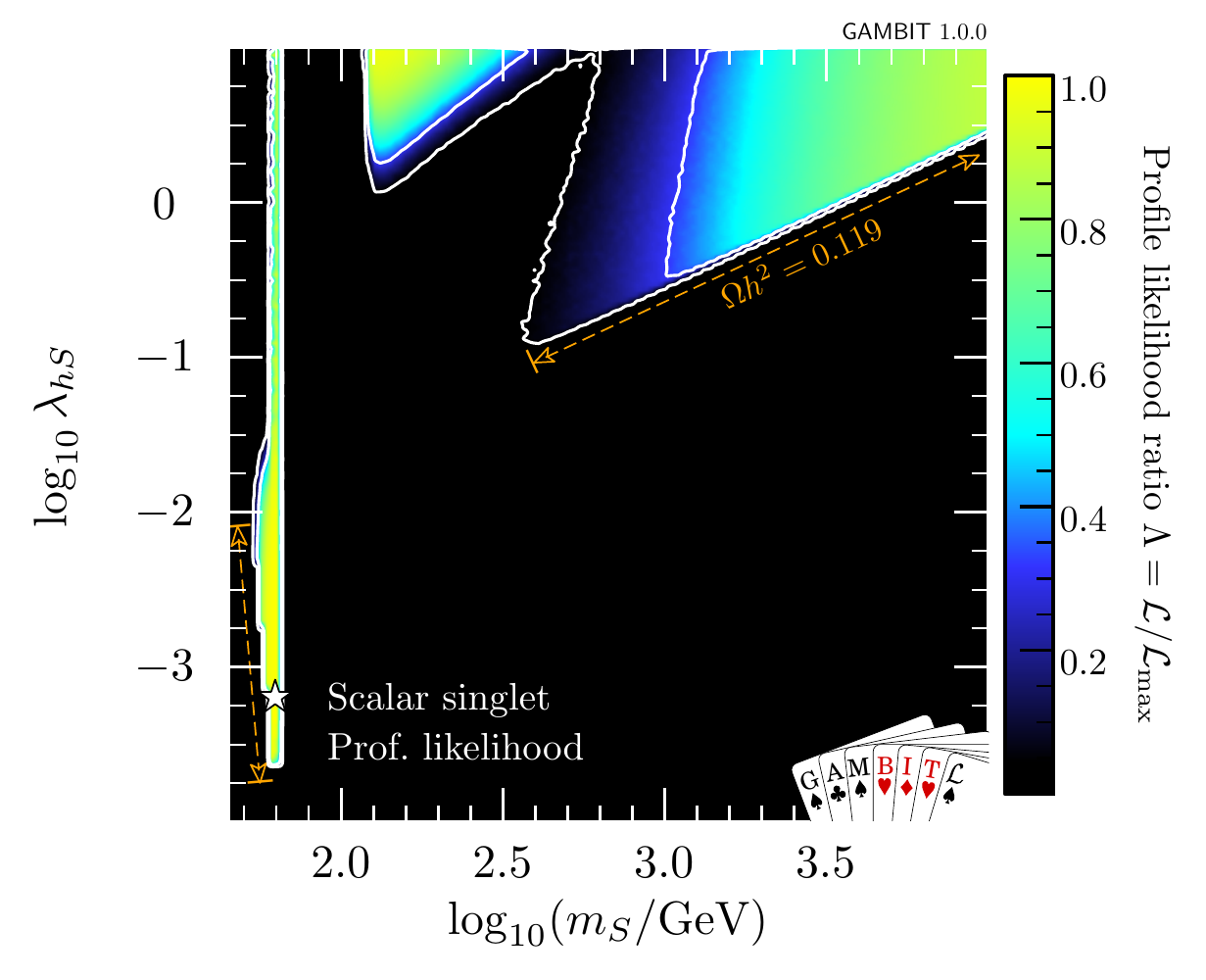}
\caption{Profile likelihoods for the scalar singlet model, in the plane of the parameters $\lhs$ and $\ms$.  The white contours indicate the $1\sigma$ and $2\sigma$ confidence regions.  The left panel is focused on the low mass resonance region, while the right shows the full parameter range.  The white star indicates the best-fit (maximum likelihood) point.  The orange lines indicate edges of the allowed region where $S$ constitutes 100\% of dark matter.  Figures from \cite{Athron:2017kgt}.}
\label{fig::Ms_lhs}
\end{figure*}

The 2D profile likelihood resulting from our global fit analysis is presented in Fig.~\ref{fig::Ms_lhs} with respect to $\lhs$ and $\ms$.  We find that the low mass resonance region, a well-known feature from previous studies, is still allowed.  However, it is heavily constrained by direct detection from lower masses, indirect detection from higher masses, Higgs invisible width from above and the relic density from below.  There also exists a narrow ``neck" region directly on the resonance, which is constrained by the Higgs invisible width from lower masses and direct detection from higher masses.  The largest allowed regions are two high-mass, high-coupling solutions.  These two regions are separated from below by the most recent LUX and PandaX direct detection exclusion limits.  The region excluded by this constraint, effectively dividing the two high mass modes, is due to the rescaling of the direct detection signals by the predicted relic density, which approximately cancels the leading $\lhs^2$-dependence of the nuclear scattering cross-section.  Therefore, since the relic density has a logarithmic dependence on $\lhs$,  for large values of this coupling the exclusion limits can reach masses of a few hundred GeV.

\begin{figure*}[h!]
\centering
\includegraphics[height=0.45\columnwidth]{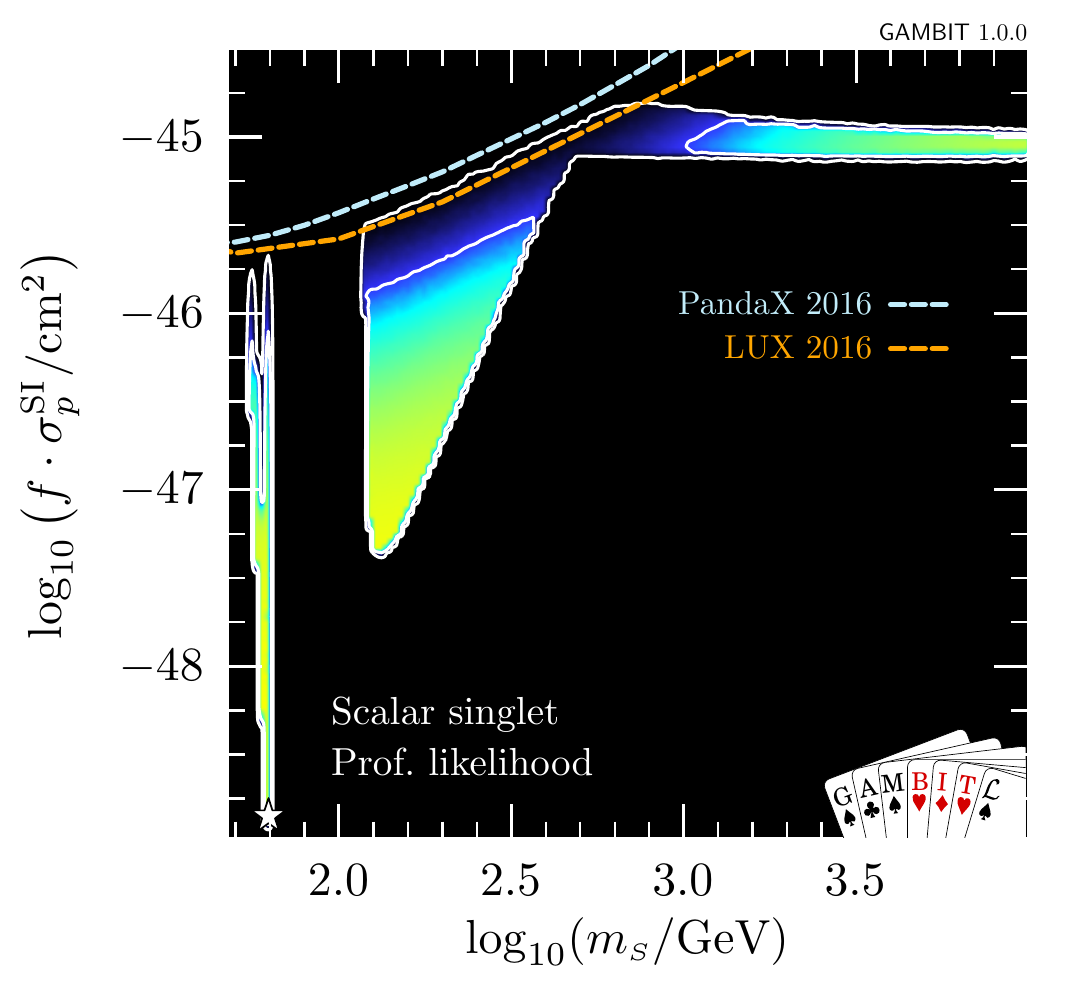}\includegraphics[height=0.45\columnwidth]{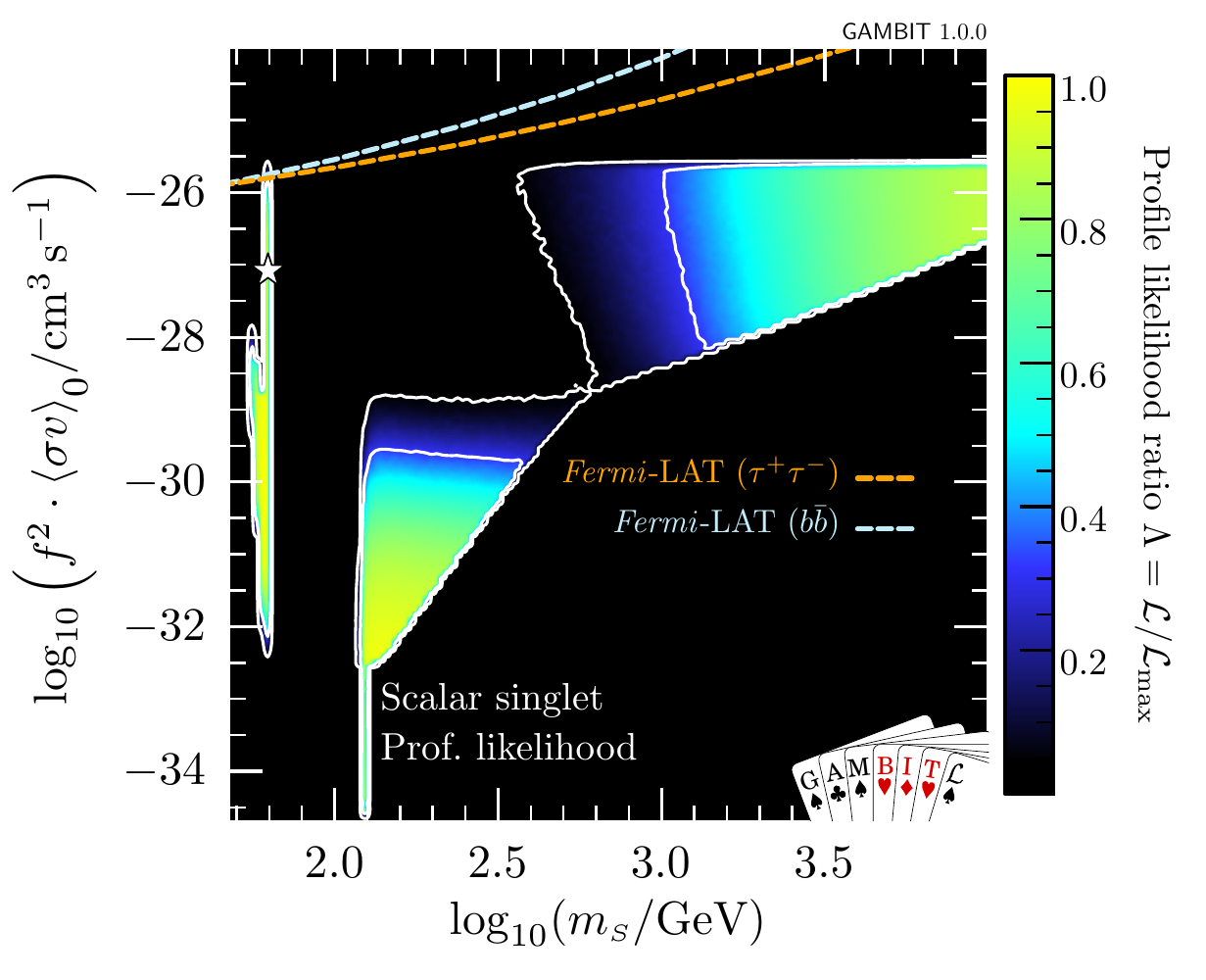}
\caption{Profile likelihoods of nuclear scattering (\textit{left}) and annihilation (\textit{right}) cross-sections for the scalar singlet model, scaled for the singlet relic abundance and plotted as a function of the singlet mass.  Here we rescale the nuclear and annihilation scattering cross-sections by $f\equiv\Omega_{\sss S} / \Omega_\text{DM}$ and $f^2$, in line with the linear and quadratic dependence, respectively, of scattering and annihilation rates on the dark matter density.  Contour lines indicate the $1\sigma$ and $2\sigma$ confidence regions.  The best-fit (maximum likelihood) point is indicated with a white star. Figures from \cite{Athron:2017kgt}.}
\label{fig::Ms_oh2_scaled}
\end{figure*}

By using the relic density as an upper limit, all points for which $\Omega_{\sss S}h^2\leq\Omega_\text{DM} h^2$ are assigned a null log-likelihood contribution and treated the same as those with $\Omega_{\sss S}h^2\ =\Omega_\text{DM} h^2$.  However, by consistently rescaling the local DM density as well as that in dwarf spheroidal galaxies, the direct and indirect detection likelihoods are not flat within this allowed region.  Thus, in contrast to pure exclusion studies, we gain additional information with some points favoured more than others.  This rescaling is clear when we present the same 2D profile likelihood with respect to cross-sections rescaled by the appropriate power of $\Omega_{\sss S} / \Omega_\text{DM}$ in Fig.~\ref{fig::Ms_oh2_scaled}, together with the experimental constraints from \textit{Fermi}-LAT, LUX and PandaX.

The best-fit point is located within the low-mass resonance region, at $\lhs = 6.5\times 10^{-4}$, $\ms=62.51$\,GeV.  This point has a combined log-likelihood of $\log(\mathcal{L})=4.566$.  This can be compared with the corresponding likelihood if each component is assigned the hypothetical `ideal' fit, which reproduces positive measurements exactly, and is equal to the background-only value for observables with only a limit.  This ideal combined likelihood is $\log(\mathcal{L})=4.673$, a difference of $\Delta\ln\mathcal{L}=0.107$ with respect to our best-fit. The best fit in the high mass, high coupling modes is at $\lhs=9.9$, $\ms=132.5$\,GeV, with $\log(\mathcal{L})=4.540$, $\Delta\ln\mathcal{L}=0.133$.  We also obtain best-fit points with the constraint that $\Omega_{\sss S}$ be within $1\sigma$ of the \textit{Planck} value.  With this constraint we find the best-fit is again in the low mass region at $\lhs=2.9 \times 10^{-4}$, $\ms = 62.27$\,GeV.  This point has $\log(\mathcal{L})=4.431$, so $\Delta\ln\mathcal{L}=0.242$ compared to the ideal model.  By making a rough estimate for the $p$-value of these points we obtain values within the range 0.4 to 0.9 -- which suggests that the fits are perfectly reasonable.  This also indicates that there is no significant preference from data for $S$ to make up either all or only a fraction of the observed DM.  See Ref. \cite{Athron:2017kgt} for a discussion regarding the interpretation of $\Delta\ln\mathcal{L}$ and the estimation of $p$-values.

From the scans using the \twalk MCMC algorithm we are also able to obtain high quality marginalised posterior distributions.  The distributions are presented in Figures 4 and 5 of Ref. \cite{Athron:2017kgt}.  We find that although the resonance region is detected, and appropriate priors are used, it is heavily penalised by its small posterior volume (less than 0.4\% of the total posterior mass), arising from the volume effect of integrating over nuisance parameters.  The penalty is sufficiently severe that this mode drops outside the $2\sigma$ credible region in the $\ms$-$\lhs$ plane, which is an indication of its heavy fine-tuning, a property naturally penalised in a Bayesian analysis.

Finally, we check vacuum stability for some interesting benchmark points.  For details of the method see Sec 4.4 of Ref. \cite{Athron:2017kgt}.  For our best-fit point, the Higgs-portal coupling $\lhs$ is too small to make a noticeable positive contribution to the running of the Higgs self-coupling, which reaches a minimum value of $-0.0375935$ at $2.523\times10^{17}$\,GeV.  The electroweak vacuum remains meta-stable for this point, with no substantial change in phenomenology compared to the SM.  Next we consider a high-mass point within our $1\sigma$ allowed region: $\lhs = 0.5$, $\ms = 1.3$\,TeV.  This point has a large enough coupling $\lhs$ that the minimum quartic Higgs coupling is positive: $0.0522133$ at $1.40006\times 10^9$ GeV.  Therefore we see that it is certainly possible to stabilise the electroweak vacuum within the singlet model whilst respecting all current constraints.

\section{Conclusion}

Scalar singlet dark matter, stabilised by a $\mathbb{Z}_2$ symmetry, is a still a phenomenologically viable model. In Ref. \cite{Athron:2017kgt}, and in the summary here, we have presented the most up to date and stringent global fit analysis, including 13 nuisance parameters.  We find that the remaining allowed regions are continually being constrained by experimental dark matter searches.  These results also serve as a validation of the new \GB software, showing that it is possible to efficiently produce high quality Bayesian and frequentist results while consistently combining many different constraints.

\section{Acknowledgements}
The work of JM was supported by an Imperial College London President's PhD scholarship.  JM gratefully acknowledges all members of the \GB collaboration, with whom this work was carried out.

\end{document}